# Nano/micro-plastics effects in agricultural landscapes: an overlooked threat to pollination, biological pest control, and food security

Dong Sheng[1,2,3,$], Siyuan Jing[1,2,4,$], Xueqing He[1,5], Alexandra-Maria Klein[6], Heinz-R. Köhler[7], Thomas C. Wanger[1,2,8,*]

1. Sustainable Agricultural Systems & Engineering Lab, School of Engineering, Westlake University, Hangzhou 310030, China
2. Key Laboratory of Coastal Environment and Resources of Zhejiang Province, School of Engineering, Westlake University, Hangzhou 310024, China
3. College of Environmental and Resource Sciences, Zhejiang University, Hangzhou 310030, China
4. Department of Environmental Science and Engineering, Fudan University, Shanghai 200438, China
5. Department of Health and Environmental Sciences, Xi'an Jiaotong-Liverpool University; Suzhou 215123, China
6. Nature Conservation and Landscape Ecology, University of Freiburg, Freiburg 79106, Germany
7. Animal Physiological Ecology, Institute of Evolution and Ecology, University of Tübingen, Tübingen 72076, Germany
8. Agroecology, University of Göttingen, Göttingen 37073, Germany

* Corresponding author: tomcwanger@gmail.com (TCW)

$ contributed equally

**Abstract**

Biodiversity-associated ecosystem services such as pollination and biocontrol may be severely affected by emerging nano/micro-plastics (NMP) pollution. We synthesized the little-explored effects of NMP on pollinators and biocontrol agents on the organismal, farm and landscape scale. For instance ingested NMP trigger organismal changes from gene expression, organ damage to behavior modifications. At the farm and landscape level, NMP will likely amplify synergistic effects with other threats such as pathogens and antibiotics, and may alter landscape properties such as floral resource distributions in high NMP concentration areas, what we call 'NMP islands'. It is essential to understand the functional exposure pathways of NMP on pollinators and biocontrol agents to comprehensively evaluate the risks for agricultural ecosystems and global food security.

**Keywords:** nano & micro plastics, pollination, biological pest control, food security, plastic pollution, synergistic threats, agricultural landscapes

# 1. Plastics, pollination, and biocontrol

Plastic pollution has been increasingly recognized as an emerging threat to human health and the environment.[1,2] The effects of microplastics (diameter ranging from 1 μm to 5 mm, hereafter MP), nanoplastics (diameter smaller than 1 μm, hereafter NP) and their associated chemicals in terrestrial ecosystems have recently moved into focus[3,4]. Publications on nano/micro-plastics (hereafter NMP) effects on the environment have increased over the last decade[5,6] (Fig. S1), showing mostly negative effects of NMP on atmosphere, biosphere, hydrosphere and pedosphere.[7,8] Previous studies have primarily focused on aquatic systems but, recently, NMP pollution in terrestrial systems has received growing attention[3] (Fig. S1). Wind, rain, and runoff facilitate NMP long-range transportation and cause plastic pollution in remote areas far away from pollution sources[9,10]. NMP have various impacts on a wide range of organisms, from microbes and plants to animals and humans[11] – for instance selectively enriching microbial communities in the 'soil plastisphere',[12] reducing Chlorophyll *b* synthesis in *Bacopa* sp.[13] and inducing oxidative stress in mice.[14] NMP also act synergistically with other threats[15] such as neonicotinoid,[16] polycyclic aromatic hydrocarbons (PAHs)[17] and toxic metals (e.g. Pb).[18] Current research mainly targets effects on single species/community, but an synthesis of NMP effects on biodiversity-associated ecosystem services such as pollination and pest control is missing,[4,19] despite these services' contribution to sustainable food production in diversified farms and landscapes.[20,21]

Anthropod pollinators are essential for the production of 70% of all globally produced food crops,[22] and biocontrol agents provide pest control services worth up to US$ 417 per ha and yr across biomes[23] with a highly favorable cost-benefit ratio of 1:250[24]. Scale effects enhance pollination and biological pest control and thereby facilitate global food security[25,26] and the effect of respective pollination and biocontrol is 32% and 23% higher in diversified than non-diversified farms[20]. At the landscape level, bee richness[27] and biocontrol agents[28] in diversified systems increased by up to four fold and 50%, respectively. However, insects as major pollinator and biocontrol agents are globally declining from habitat loss, pathogens and parasites, climate change, and the overuse of pesticides[29,30]. Pollinators and biocontrol agents are likely exposed to and affected by NMP in similar ways to other terrestrial and aquatic organisms[31,32]. For instance, NMP may act synergistically with other threats[15,17] to pollinators

and biocontrol agents[33,34] because MP can act as carriers and releasers of pollutants and then facilitate organismal ingestion[16,35]. Moreover, plastic pollution for instance from plastic mulching can change the soil structure and properties,[36,37] with implications for plant growths and floral resource distributions in agricultural landscapes.[38,39] However, a synthesis of all known direct and indirect effects of NMP on pollination and biological pest control at the organismal, farm and landscape scale is missing but urgently needed to guide policies and future research activities.

We use a systematic review to quantify all known and potential NMP exposure pathways, as well as direct and indirect effects of NMP on pollinators and biocontrol agents (Fig. 1). We focus on NMP effects individually and in synergy with other threats from the organism to farm and the landscape scale. After highlighting important research gaps, we close with a research agenda to avoid potentially severe, yet unrecognized threats to global food production.

## 2. Systematic literature review

We use a systematic review to understand the effects of NMP on pollination services and biological pest control (Fig. S2, Fig. S3). For pollination services, we searched the Web of Science on February 13th, 2024 with the search string "TS = ((nanoplastic* OR microplastic*) AND (pollinat* OR (bee OR bees) OR honeybee*))". We found 21 studies out of which 16 were included in our review. Nine research articles reported experiments with NMP on honeybees, five of which focused on NMP effects[40–44] and the other three also considered combined effects with other substances.[45–47] Three studies confirmed honeybee's environmental exposure to NMP.[46,48,49] The transfer of NMP within bee hives and its threat to honey products were also investigated.[48–50] The remaining five papers were non-quantitative summaries.[4,19,51–53]

For NMP effects on biological pest control, we used the search string "TS = ((nanoplastic* OR microplastic*) AND ("biological pest control" OR "biological control" OR pest OR pests OR pest-control OR "control agent*"))" in Web of Science on February 16th, 2024. Of the 22 studies listed, only three were relevant for pest control in agriculture[54–56] indicating that the topic is largely unexplored. For a summary of all identified effects, please see Tabel 1.

## 3. NMP exposure pathways to pollinators and biocontrol agents

Pollinators and biocontrol agents are at risk from a plethora of NMP exposure pathways. Plastics accumulate in agricultural landscapes up to ~$2\times10^3$ particles/kg in farm soils[37,57]. Direct sources of macroscopic plastic particles include plastic mulch films[37,58], and protective nets[59] that result in NMP due to photodegradation, mechanical abration, and biodegradation [60,61]. Indirect plastic inputs result from NMP-polluted sewage sludge,[62,63] fertilizers,[64] compost,[37] irrigation,[65] and manure.[66] Different kinds of NMP are present as suspended particulates in the atmosphere, as atmospheric fallout, road dust, and even as a substantial component of particulate matter $PM_{2.5}$ [67] (Fig. 1). Suspended airborne NMP may attach to insects' surfaces, and the deposited particles found in water and on the inflorescences of flowering plants[68] may enter insects through ingestion. Additionally, NMP in agricultural soils[69] may threaten ground-nesting and soil-nesting bees. Pollinators ingest NMP[46] or collect plastic as nesting materials,[70] and then transfer them into their nests and larvae[50]. Certain bee keeping practices also directly introduce NMP into the nest.[48] All of the above indicates direct exposure of pollinators to NMP pollution.

Biocontrol agents are also likely to ingest NMP while foraging. Currently, it is difficult to assess whether an increased NMP exposure occurs via bioaccumulation over trophic levels (biomagnification) and, hence, the potential bioaccumulation risk by NMP to pest control agents. This is because general intracellular uptake of plastics is limited to sizes below 1 µm,[71] which are difficult to quantify analytically[72] (but see Anbumani and Kakkar[73]). Overall, exposure studies are needed to vastly expand our understanding of the complex NMP exposure pathways on pollinators and biocontrol agents on the organismal but even more so on the farm and landscape level.

## 4. Direct microplastic effects at the organismal level

### 4.1. Pollinators

Current evidence from laboratory experiments suggests that MP have limited lethal effects on honeybees in general,[40–42] but reduced the survival rate of newly-emerged worker bees.[45] MP from food and water resources accumulated (i.e., continuous addition to the intestinal lumen without complete remorval) in bees' digestive system, especially in the midgut and hindgut.[42,45–47] Accumulated MP in bee guts can harm tissues (5, 50 µm polystyrene PS-MP)[46], induce intestinal dysplasia (1, 100 µm PS-MP)[45], and alter gut microbiota composition (1, 25, 100 µm PS-MP)[45,47]. Small-sized MP (down to 5 µm) could also enter the respiratory system and accumulate in tracheae,[46] or penetrate and accumulate in the brain.[44] The circulatory system is affected by early exposure to MP, with a shift of plasmatocytes and prohemocytes (0.5 mm PS-MS, 0.6 mm polyethylene terephthalate (PET-MP).[43] MP also stimulate gene expressions related to oxidative stress, the immune system, and detoxification.[45,47]

MP exposure led to a series of behavioral changes. For instance, polyethylene (PE)-MP intake led to altered food consumption (0.1 – 100 mg/L, 0.2 – 9.9 µm) and caused inconsistent proboscis extension responses (hereafter PER) in honeybees.[41] Similarly, Pasquini et al.[44] reported reduced sucrose responsiveness and impaired learning and memory after PS-MP treatment (0.5 – 50 mg/L, 4.8 - 5.8 µm). Moreover, exposure to PS-MP (1 – 100 mg/L, 27 and 93 µm)[40] and polyester (PLY)-MP (10 – 100 mg/L, aerodynamic diameter[74] = 84 µm)[42] led to reduced food intake (but see Balzani et al.[41]). Early exposure to PET-MP (12.5 mg/L, 0.6 mm) changed locomotion behaviors of adult bees, including more resting and more interactions.[43] As honeybees showed no preference or avoidance between food and water resources with or without PLY-MP (100 mg/L, 84 µm),[42] bees may not have the ability to distinguish and, hence, avoid MP in the real environment. Current MP toxicity studies focus on honeybees only and, hence, the potentially different physiological and behavioural effects across different pollinator groups and in different life stages[50,75] must be urgently addressed to understand the implications for general pollination services.[76]

**4.2. Pests and biocontrol agents**

Research on the effects of MP on pests and biocontrol agents is in its infancy, with only three laboratory studies available. Rondoni et al.[55] found that PE-MP (5% of the soil weight, 157 µm) exposure reduces preference for oviposition on plant leaves in black fungus gnats (*Bradysia difformis*, an important crop pest). Thormeyer and Tseng found no fitness-related effects of PS-MP (200 – 20k items/mL, 4.8 – 5.8 µm) on *Culex pipiens* and *Culex tarsalis* larvae.[56] For biocontrol agents, Pazmiño et al.[54] reported inconclusive evidence that MP (5% of feed, 2.12 mm polylactic acid (PLA), 1.71 mm PE, or 1.06 mm PS) exposure could affect the larval development of *Hermetia illucens*, a pest control agent for filth flies[77]. These studies confirm initial effects of MP exposure on agricultural pests and biocontrol agents. MP biomagnification, is documented in marine systems,[78] may lead to increased MP exposure for pest predators and stronger impacts on their biological pest control services. Baseline research is urgently needed on direct MP effects on pests and biocontrol agents.

## 5. Direct nanoplastic effects on pollinators and biocontrol agents at the organismal level

NP exhibit novel toxicity effects due to their distinct physical, chemical, and biological properties, such as their shapes and protein/eco-corona.[79] NP can not only cross some biological barriers and act as carriers of toxicants, but also modulate organismal functions such as growth and oxidative cell stress.[80,81] We found only four studies that involved direct NP effects on pollinators.[41,45,46,53] A review of NMP effects on pollinators focuses broadly on physiological aspects, but not on the differences between MP and NP uptake pathways and tissue translocation.[53] Similarly to MP, Wang et al.[45] showed that oral exposure of NP (100 nm) significantly reduced body weight and survival rate, and induced intestinal dysplasia in honeybees. Deng et al.[46] found that NP (0.5 μm) was especially harmful (compared with MP) to honeybees by accumulating in the midgut and trachea tissues, and stimulating gene expression. Balzani et al.[41] used particles between 0.2 – 9.9 μm, suggesting that their findings on changes in feeding behaviors and mortality might be combined effects of MP and NP (Fig. S4). Current studies indicate that smaller size NP tend to have stronger negative effects probably due to increased ability to penetrate biological barriers. Although similar effects to pollinators may be expected, our literature search did not reveal any publications of NP effect on pests and biocontrol agents.

Research is urgently needed to better understand the effects of both MP and NP on pollinators, but even more so on biocontrol agents and pests. This is, because mechanistically, NP toxicity is complex and often linked to MP exposure. For instance, NP related oxidative cell stress[80] can lead to DNA damage, apoptosis, and cell death.[82] Such effects may impair pollinators' memory, learning, and other behaviors such as reduced reproductive success with implications for pollination services.[83,84] Moreover, the characteristics of nanoparticles suggest different mechanisms between NP and MP, but the available studies for pollinators and pest control agents show no fundamentally different effect from MP and NP exposure. This may be, because the NMP used in current studies vary greatly in terms of doses, shapes (spheres,[41] fibers,[42] and fragments[40]), diameters and chemical components (Fig. S4), which will mediate organismal effects. Lastly, NP toxicity may be strongly related to MP exposure level and degradation rate, both at the organismal and lndscape level. For example, gut bacteria in can degrade MP particles in honey bee hindguts[45] and thereby exacerbate NP exposure. In agricultural landscapes, MP degradation for instance in soils will increase plant, pollinator and pest control agent exposure to NP, eventually affecting food production and security to an unknown extend.

## 6. Indirect effects of NMP in agricultural landscapes

Agricultural landscape complexity not only mediates pollination and biological pest control services[85,86] but also affects direct and indirect NMP deposition. Plastic accumulation and retention is likely to be driven by farm and landscape features.[87] For instance, individual trees and hedge rows at the farm level will block runoff and retain plastic particles. At the landscape scale, forests and semi-natural habitats can capture fine particulate plastic for instance contained in aerosols.[8,88] The spatial configuration of these structures should correspond to areas with buildup of high NMP concentrations that mediate plastic concentration in natural landscapes,[89] which we refer to as "NMP islands".

### 6.1. NMP may amplify other threats

NMP islands in agricultural landscapes may amplify other environmental threats to pollinators and pest control agents such as chemical pollution[90,91] and pathogens.[92] For example, the interaction toxicity of NMP with pesticides[93] is determined by their physicochemical characteristics such as plastic type, size,[94] surface charge,[95] and concentration,[96,97] which brings indirect risk for terrestrial organisms.[67] Pollinators are known to be affected by pesticides such as neonicotinoids[98] but also by fungicides[99] for which exposure may be modified through the catalytic activity of NMP. For instance, the survival rate of honeybees dramatically decreased in a combination-treatment with tetracycline and MP as opposed to individual treatments.[47] Mechanistically, NMP can adsorb substances such as PAHs, or persistant organic pollutants (POPs).[100] This may lead to the accumulation of toxic chemicals on NMP surfaces and potentially a modification of interaction toxicity and overall higher substance concentrations. In the marine environment, adsorbed contaminants on MP surfaces increased toxicity towards mussel embryo.[101] In other examples, however, very high MP concentrations reduced the effectiveness of thiacloprid in chironomid larvae, possibly by diminishing the uptake of the pesticide in the gastrointestinal tract.[102] As pesticides affect both pollinators and biocontrol agents, which in turn are heavily influenced by farm and landscape level effects,[103,104] NMP islands may likely mediate these relationships further. Future research should investigate interaction toxicity mechanisms and effects on ecosystem service providers in particular with well known threats from neonicotinoid pesticides across spatial scales.

Pollinators and biological pest control agents are devastated by pathogenic viruses and bacteria[105,106] and depend on multi-taxa interactions ranging from invertebrates to microorganisms and fungi.[107,108] In pollination, the impacts of viruses on their hosts are exacerbated by other major stressors such as parasites, poor nutrition, and exposure to chemicals.[109,110] NMP can further enhance the invasion of Israeli Acute Paralysis Virus and *Hafnia alvei* to honeybees by affecting cell membranes (especially NP) and immune systems.[45,46] Moreover, microbial communities can colonize plastic particles, which may facilitate spreading pathogenic bacteria and fungi, while becoming reservoirs for antibiotic and metal resistance genes in soils.[12] Overall, NMP islands facilitate may faciliate unintentional interaction toxicity and higher susceptibility to established and new pathogens in agricultural landscapes, of which most mechanisms and implications urgently require more research.

### 6.2. NMP may alter agricultural landscapes and ecosystem services

NMP islands may also indirectly affect pollinators and biocontrol agents through changes in agricultural landscapes. It is now well understood how farm and landscape level diversification affects biodiversity-mediated ecosystem services like pollination and biological pest control.[85,86,111] For instance, distribution and amount of semi-natural habitats modify abundance and diversity of pollinators and biocontrol agents.[112,113] Mechanistically, patterns are driven by floral resources, nesting opportunities, and chemical inputs,[114] which may be modified from NMP effects on soil properties, plant growth, and plant communities[38,115] as well as microbial communities.[116] The mixed NMP effects on plant growth and yield are highly species-dependent,[117] which can affect plant productivity, community structure.[118,119] Moreover, diverse floral resources can mitigate neonicotinoid and fungicide impacts on wild pollinators.[99,120] Hence, a reduction in floral resources in NMP islands may affect colony survival and increase exposure to pesticides or other agricultural chemicals.[99] All of these effects are unstudied but are highly likely to modify the resources available in plant-animal interactions and hence, effectiveness of pollination and biocontrol services.

## 7. Implications on food security and a way forward

The world is already facing massive impacts on food security by climate change, pests and diseases affecting yields, and conflicts preventing access to safe and nutritious food.[121] The above effects of NMP on pollination and biocontrol services may further exacerbate food insecurity. At organismal scale, current evidence suggests that service providers experience sublethal yet profound physiological changes[40–42,54] that are likely exacerbated with other stressors.[45–47,100–102] At the farm and landscape scale, changes in resource availability (e.g., amount or species assemblages of host plants)[118,119] or plant-soil system characteristics[38,115] may restrict distributions of pollinators and biocontrol agents. At NMP islands, pollination and pest control service efficiency may be further reduced, which may alter climate effects on crop distributions and yields. Moreover, a diverse diet requires various species to provide pollination and biocontrol services for a broad range of crop species.[122] Species and variety-dependent NMP effects on pollinators and pest control agents may, therefore, constrain the choice of crop species and varieties in the future. Lastly, the projected surge of plastic waste accumulation (12 million metrictons by 2050) and NMP pollution in the coming decades[1,123] adds to the current risks of food insecurity,[124] and threatens the stability of global food production. In addition to acknowledging plastic pollution as a key concern for biodiversity and associated services,[125] we advocate for research on how diversified agricultural landscapes[126] mediate the tradeoff between pollinator and pest control benefits and accumulation effects at "NMP islands" to insure long term maintenance of crop yields and food security.[127]

Future research should target the development and refinement of methods that can be applied in laboratory, semi-field, and field studies to address global food security implications.

1. **Methods to detect NMP in environmental samples.** Despite more effective methods for NMP detection in environmental samples are continuously developed,[72,128] the detection of small-sized plastics (< 1 µm) in real environments and in organisms remains a critical bottleneck. Moreover, methods to automate sample preparation and analysis are currently a bottleneck in NMP research especially in terrestrial systems, where samples are comprised of an organic and much more complicated matrix than in water.[128,129] In addition, it becomes increasingly clear that plastic pollution exhibits less acute and more chronic effects, which require standard methods to effectively track NP and their interaction effects in pollinators, biocontrol agents, and more broadly ecosystem service providing insect communities.

2. **Ecosystem Ecotoxicology of NMP.** Laboratory studies have shown the ecotoxicity of NMP on different organisms, but only to a limited extent on pollinators and biocontrol agents. In addition, concentrations used in current studies are often likely too high compared to largely unknown real-field NMP exposure, similar to the aquatic environment.[71] More systematic perspectives involving various ecosystem agents should be adopted, for instance the "novel epidemiology" concept,[130] whereby a plant-pollinator-pathegon network is used to analyze plant-pollinator extinction. As NMP properties change greatly due to weathering and chemical degradation, different NMP properties across realistic environmental concentrations must be investigated on commercial (e.g., honey bees) and wild pollinators (e.g., solitary bees and hoverflies acting as pollinators and pest-control agents) in controlled environmental conditions. In addition, the newly developed methods above should identify realistic NMP concentrations to be used in semi-field and field effect studies on pollinators and biocontrol agents' acute and chronic lethal and sublethal toxicity and the effects on behavior and their ecosystem services on the semi-field to field scale. Specifically, NMP metabolites, leachate, and interaction toxicity require more attention.

3. **NMP impact mitigation.** In urban environments, $PM_{2.5}$ contains up to 15% fine particulate plastic and can be mitigated through green wall structures and planted roofs, which would also reduce the exposure of pollinators. Moreover, small trees and shrubs can also improve air quality in streets.[131,132] Designing vegetation barriers for NMP transfer depends on the choice of plant species and composition, and their spatial configuration.[133,134] This could be integrated into systematic conservation planning[135] practices to conserve endangered pollinators and biocontrol agents. However, understanding the role of landscape heterogeneity in potentially mitigating plastic but also chemical pollution in agricultural landscapes across scales is an emerging area of research.

Additional funding should be allocated specifically to understand NMP effects across scales on biodiversity associated ecosystem services such pollination and biological pest control in pursuit of the Global Biodiversity Framework's roadmap for biodiversity conservation[125] and a food secure future.

79. Junaid, M. & Wang, J. Interaction of nanoplastics with extracellular polymeric substances (EPS) in the aquatic environment: A special reference to eco-corona formation and associated impacts. *Water Res.* **201**, 117319 (2021).

80. Sun, X. *et al.* Toxicities of polystyrene nano- and microplastics toward marine bacterium Halomonas alkaliphila. *Sci. Total Environ.* **642**, 1378–1385 (2018).

81. Jiang, W., Kim, B. Y. S., Rutka, J. T. & Chan, W. C. W. Nanoparticle-mediated cellular response is size-dependent. *Nat. Nanotechnol.* **3**, 145–150 (2008).

82. Prüst, M., Meijer, J. & Westerink, R. H. S. The plastic brain: neurotoxicity of micro- and nanoplastics. *Part. Fibre Toxicol.* **17**, 24 (2020).

83. Farooqui, T. Iron-induced oxidative stress modulates olfactory learning and memory in honeybees. *Behav. Neurosci.* **122**, 433–447 (2008).

84. Massaad, C. A. & Klann, E. Reactive oxygen species in the regulation of synaptic plasticity and memory. *Antioxid. Redox Signal.* **14**, 2013–2054 (2011).

85. Haan, N. L., Zhang, Y. & Landis, D. A. Predicting Landscape Configuration Effects on Agricultural Pest Suppression. *Trends Ecol. Evol.* **35**, 175–186 (2020).

86. Toledo-Hernández, M. *et al.* Landscape and farm-level management for conservation of potential pollinators in Indonesian cocoa agroforests. *Biol. Conserv.* **257**, 109106 (2021).

87. Dikareva, N. & Simon, K. S. Microplastic pollution in streams spanning an urbanisation gradient. *Environ. Pollut.* **250**, 292–299 (2019).

88. Allen, S. *et al.* Atmospheric transport and deposition of microplastics in a remote mountain catchment. *Nat. Geosci.* **12**, 339–344 (2019).

89. Lwanga, E. H. *et al.* Microplastic appraisal of soil, water, ditch sediment and airborne dust: The case of agricultural systems. *Environ. Pollut.* **316**, 120513 (2023).

90. Almeida, R. A., Lemmens, P., De Meester, L. & Brans, K. I. Differential local genetic adaptation to pesticide use in organic and conventional agriculture in an aquatic non-target species. *Proc. R. Soc. B Biol. Sci.* **288**, 20211903 (2021).

91. Vanbergen, A. J. A cocktail of pesticides, parasites and hunger leaves bees down and out. *Nature* **596**, 351–352 (2021).

92. Vurro, M., Bonciani, B. & Vannacci, G. Emerging infectious diseases of crop plants in developing countries: impact on agriculture and socio-economic consequences. *Food*

**Acknowledgements**

TCW was funded by a Westlake University Start-up fund.

**Author Contributions**

Conceptualization: XH, TCW;

Data collection: DS, SJ;

Methodology and Analyses: DS, SJ, XH, TCW;

Visualization: DS, SJ, TCW;

Writing – original draft: DS, SJ, XH, TCW;

Writing – review & editing: DS, SJ, XH, HRK, AMK, TCW;

Funding acquisition: TCW;

Project administration: TCW;

Supervision: TCW;

Figures

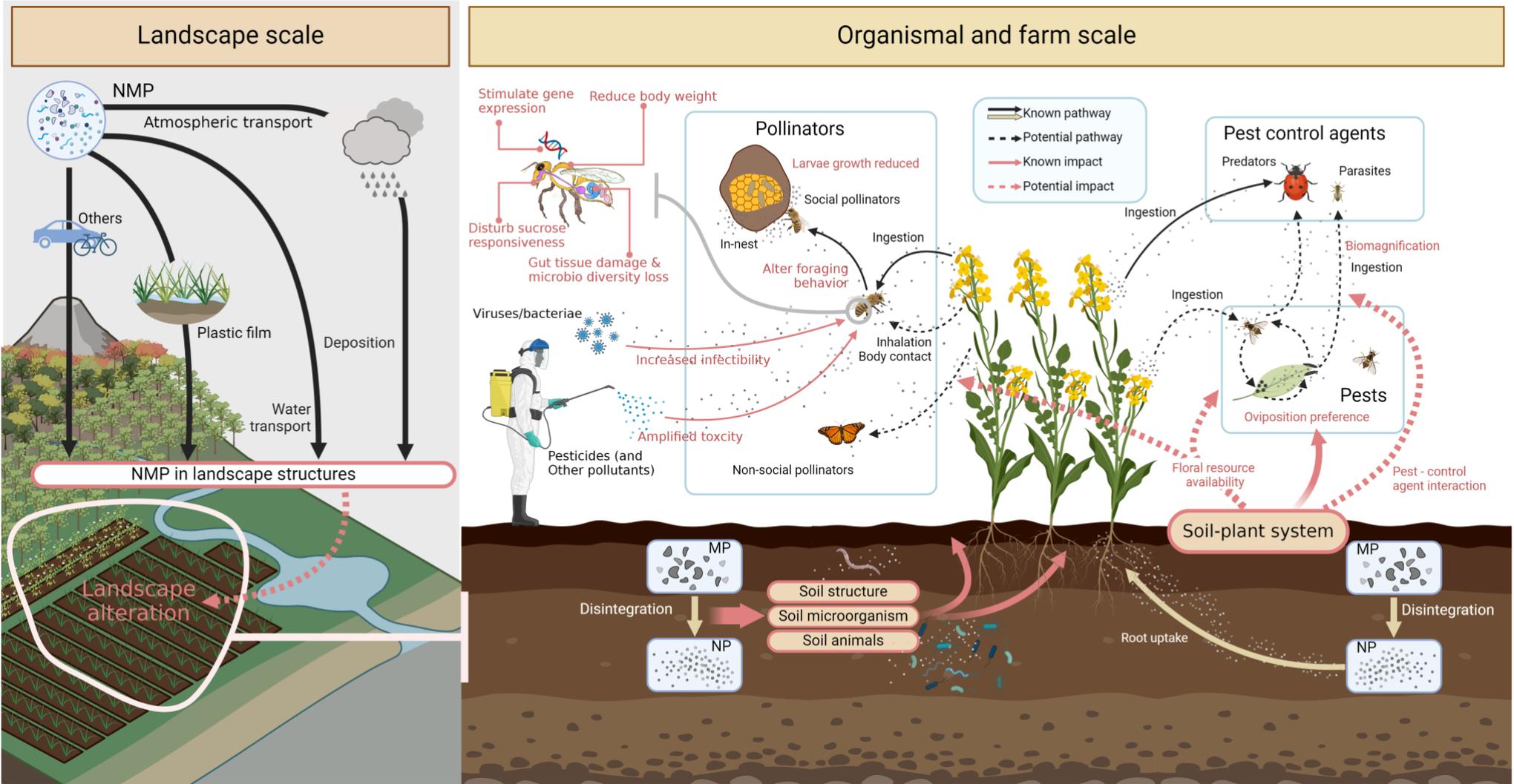

**Figure 1.** Nano/micro-plastic (NMP) exposure pathways and direct and indirect effects on pollinators, pests and pest control agents at the landscape and farm scale. Known or evidence-supported pathway and effect were showed with solid arrows and anticipated with dashed arrows. The relative importance of these pathways and effect is likely going to differ depending on microplastic types and characteristics such as size and shape. For a detailed description see the main text. Abbreviations: MP = microplastics, NP = nanoplastics, NMP = nano- and microplastics.

7 **Tables**

8 **Table 1.** Known impacts of NMP on pollinators, pests and biocontrol agents.

| Target | Type | Impact | Aspect | Effect | Evidence | References |
|---|---|---|---|---|---|---|
| Pollinator *(Apis mellifera)* | MP | Direct | Mortality | No effect or low effect on survival rate | ** | 40–42,47,50 |
| | | | Biomass | Reduce body weight | ? | 40,47 |
| | | | Food consumption | Less intake of sucrose solution | ? | 40–42 |
| | | | Behaviors | Disturb sucrose responsiveness | ** | 41,44 |
| | | | | No effect to sucrose preference | ** | 41,42 |
| | | | | Impair learning and memory | * | 44 |
| | | | Alimentary system | Damage midgut tissue | * | 46 |
| | | | | Gut microbiota diversity loss | * | 47 |
| | | | Gene expression | Stimulate expressions related to immunity, detoxification, etc. | ** | 46,47 |
| | | Indirect | Infection | More infectious to viruses | * | 46 |
| | | | Amplifier of pollutants | More vulnerable to antibiotics | * | 47 |
| | NP | Direct | Mortality | Reduce survival rate | ** | 41,45 |
| | | | Biomass | Reduce body weight | * | 45 |
| | | | Food consumption | More intake of food | - | 41 |
| | | | Behaviors | Disturb PER to sucrose | - | 41 |
| | | | | No effect to sucrose preference | - | 41 |
| | | | Alimentary system | Damage midgut tissue (stronger than MP) | * | 46 |
| | | | | Induce intestinal dysplasia | * | 45 |
| | | | | Certain gut microbiota loss | * | 45 |
| | | | Gene expression | Stimulate expressions related to immunity, detoxification, etc. | ** | 45,46 |
| | | Indirect | Infection | More infectious to bacteria | * | 45 |
| Pollinator *(Apis cerana)* | MP | Direct | Alimentary system | Damage midgut tissue | * | 46 |
| | | | Gene expression | Alter gene expressions | * | 46 |
| | | Indirect | Infection | More infectious to viruses | * | 46 |
| | NP | Direct | Alimentary system | Damage midgut tissue (stronger than MP) | * | 46 |
| | | | Gene expression | Alter gene expressions | * | 46 |
| Pollinator *(Partamona helleri)* | MP | | Biomass | Increase body weight | * | 43 |
| | | | Circulatory System | Change hemocyte counts | | |
| | | | Foraging behavior | Disturb walking behavior | * | 43 |
| Biocontrol agent *(Hermetia illucens)* | MP | Direct | Larvae | Alter larvae biomass | - | 54 |
| Pest *(Bradysia difformis)* | MP | Indirect | Oviposition | Lower oviposition interest to polluted plant-soil systems | * | 55 |
| Pest *(Culex pipiens & Cx tarsalis)* | MP | Direct | Fitness-related | No effects on body size, development and growth rate | * | 56 |

10  **  Some supporting literature, i.e., ≥ 2 supporting studies;

11  *   Limited supporting literature, i.e., < 2 supporting studies;

12  -   Results not obvious, i.e., the only supporting study used mixtures of NP and MP, or showed mixed effects within same study;

13  ?   Controversial, where conflicting results were provided by different studies.

**SUPPLEMENTARY MATERIAL**

**Nano/micro-plastics effects in agricultural landscapes: an overlooked threat to pollination, biological pest control, and food security**

Dong Sheng[1,2,3,$], Siyuan Jing[1,2,4,$], Xueqing He[1,5], Alexandra-Maria Klein[6], Heinz-R. Köhler[7], Thomas C. Wanger[1,2,8,*]

1. Sustainable Agricultural Systems & Engineering Lab, School of Engineering, Westlake University, Hangzhou 310030, China
2. Key Laboratory of Coastal Environment and Resources of Zhejiang Province, School of Engineering, Westlake University, Hangzhou 310024, China
3. College of Environmental and Resource Sciences, Zhejiang University, Hangzhou 310030, China
4. Department of Environmental Science and Engineering, Fudan University, Shanghai 200438, China
5. Department of Health and Environmental Sciences, Xi'an Jiaotong-Liverpool University; Suzhou 215123, China
6. Nature Conservation and Landscape Ecology, University of Freiburg, Freiburg 79106, Germany
7. Animal Physiological Ecology, Institute of Evolution and Ecology, University of Tübingen, Tübingen 72076, Germany
8. Agroecology, University of Göttingen, Göttingen 37073, Germany

* Corresponding author: tomcwanger@gmail.com (TCW)
$ contributed equally

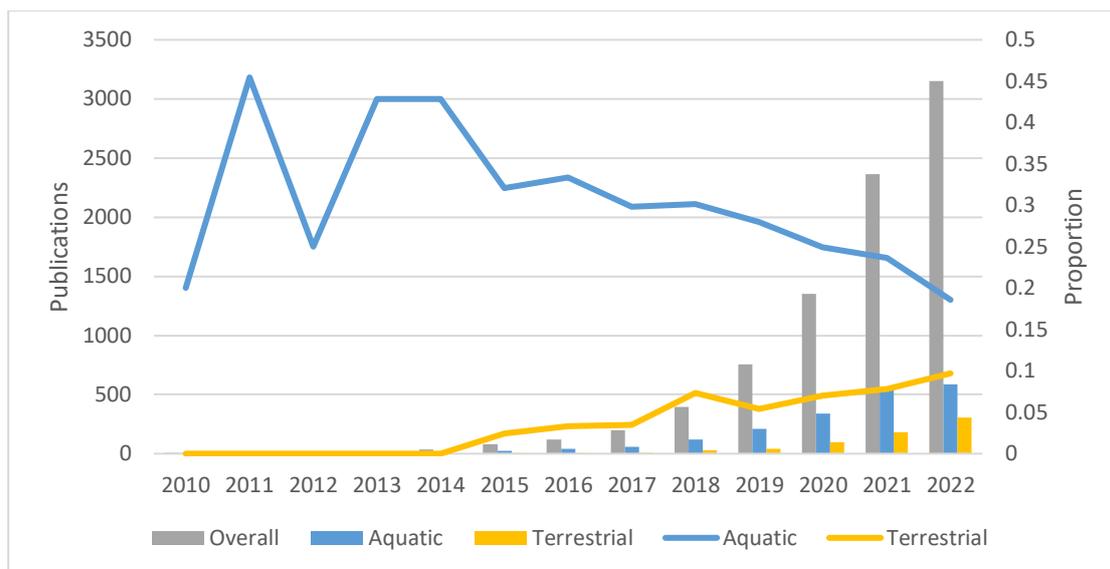

**Figure S1.** The trends of publications about NMP since 2010s. Grey/blue/yellow bars represent number of publications on overall, aquatic, and terrestrial NMP studies, respectively. Blue/yellow lines depict the proportion of aquatic/terrestrial system-related NMP studies. This figure is based on a Web of Science search on June 26th, 2023 with the following search string: "AK = (microplastic* OR nanoplastic*)", "AK = ((microplastic* OR nanoplastic*) AND (aquati* OR marine* OR water* OR river* OR lake*))", "AK = ((microplastic* OR nanoplastic*) AND (terre* OR edaph* OR soil* OR land* OR field*))", respectively.

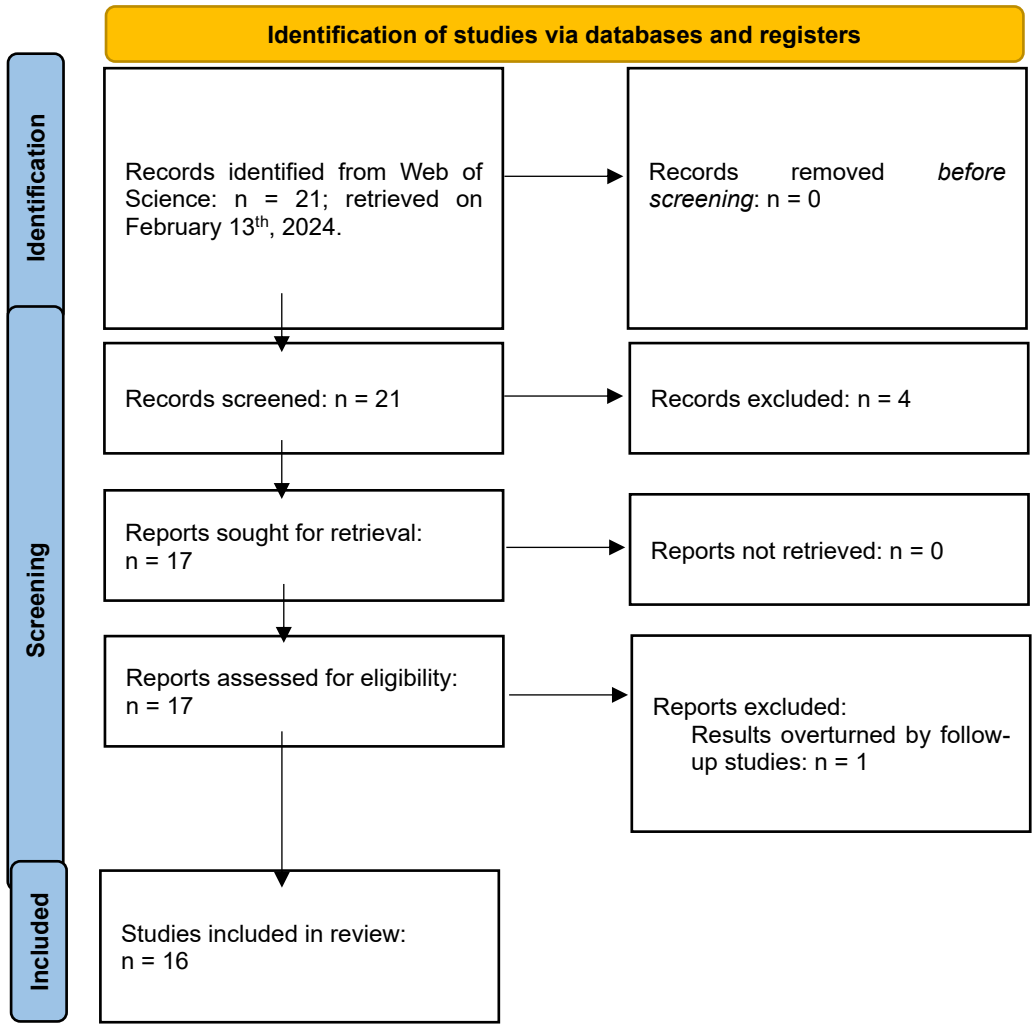

**Figure S2.** The PRISMA Chart of the systematic literature review Nano/Microplastic effects on pollination services.

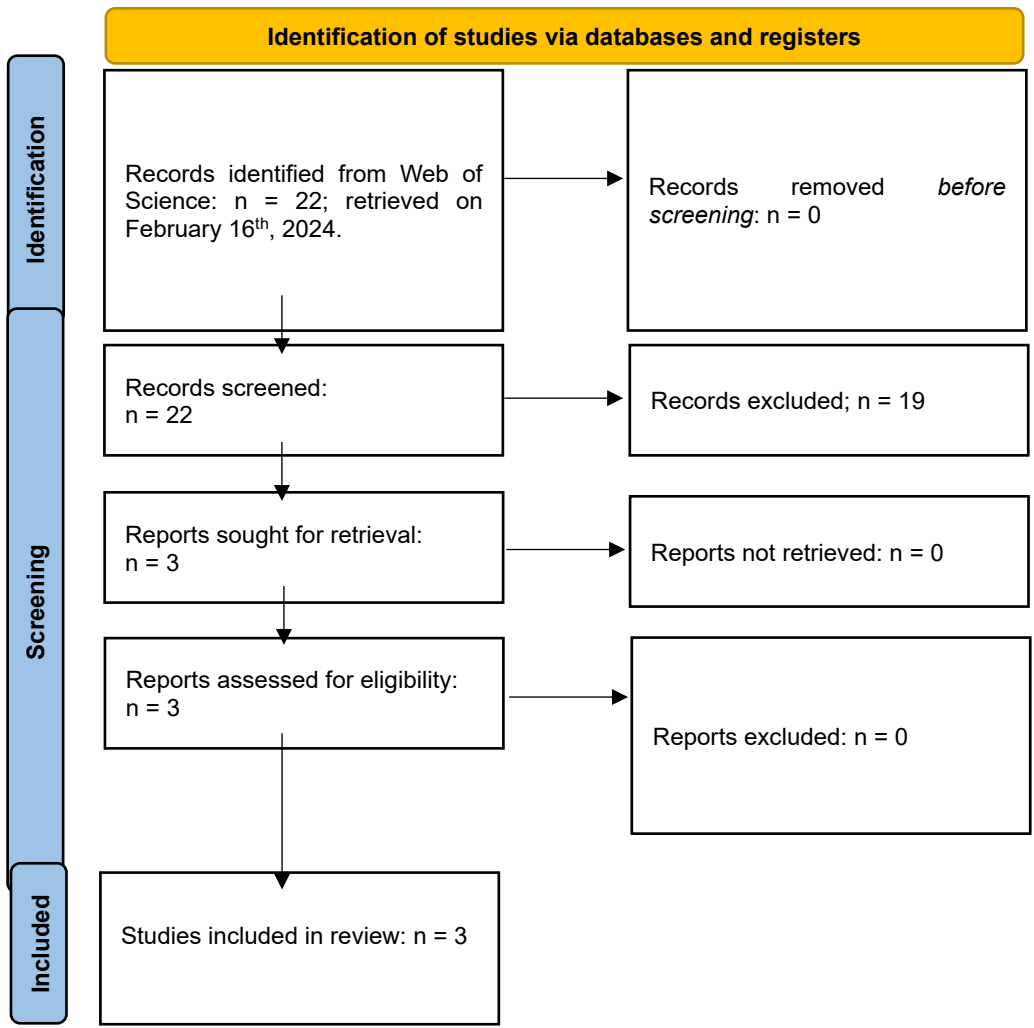

**Figure S3.** The PRISMA Chart of the systematic literature review Nano/Microplastic effects on biological pest control services.

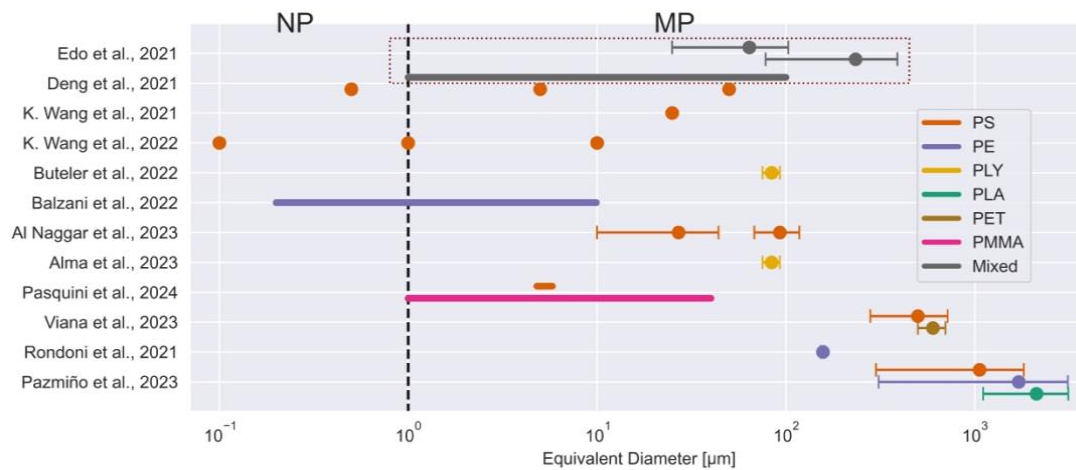

**Figure S4:** Equivalent diameters of NMP from literature, including field investigations and experiment setups. Scatters with error bars represent mean diameters ± standard errors, and lines represent diameter spans (minima ~ maxima). Equivalent diameters of fiber-shaped particles were calculated as aerodynamic diameters.[74] Overall, current experiments varied in the NMP types, diameters, and doses across lab and field experiments. Thus, with the current available studies, it is difficult to understand real implications of plastic pollution on pollinators and pest control agents. Abbreviations: NP = nanoplastics, MP = microplastics, PE = polyethylene, PS = polystyrene, PLY = polyester, PLA = polylactic acid, PET = polyethylene terephthalate, PMMA = polymethyl methacrylate.